\documentclass[preprint,aps,12pt,showpacs,nofootinbib,tightenlines]{revtex4}
\usepackage{mathrsfs}
\usepackage{amsmath}
\usepackage{amssymb}
\usepackage{epsfig}
\usepackage{graphicx}
\newcommand{\psl}{ P \hspace{-2.4truemm}/}
\newcommand{\esl}{ \epsilon \hspace{-2.1truemm}/ }

\newcommand{\pvsl}{ p \hspace{-2.0truemm}/_{K^*} }
\def\be{\begin{eqnarray}}
\def\en{\end{eqnarray}}
\def\non{\nonumber\\}

\def\ra{\rangle}

\def\sl{\!\!\!\slash}
\def\prd{{Phys. Rev. D}~}
\def\prl{{ Phys. Rev. Lett.}~}
\def\plb{{ Phys. Lett. B}~}

\def\epjc{{ Eur. Phys. J. C}~}
\newcommand{\acp}{{\cal A}_{CP}}
\begin{document}
\title{Study of scalar meson $a_0(1450)$ from $B \to a_0(1450)K^*$ Decays }
\author{Zhi-Qing Zhang
\footnote{Electronic address: zhangzhiqing@haut.edu.cn} } 
\affiliation{\it \small  Department of Physics, Henan University of
Technology, Zhengzhou, Henan 450052, P.R.China } 
\date{\today}
\begin{abstract}
In the two-quark model
supposition for the meson $a_0(1450)$, which can be viewed as either the first excited state (scenario I) or the lowest lying state (scenario II),
the branching ratios and the direct CP-violating
asymmetries for decays $B^-\to  a^{0}_0(1450)K^{*-}, a^{-}_0(1450)K^{*0}$ and $\bar B^0\to  a^{+}_0(1450)K^{*-}, a^{0}_0(1450)\bar K^{*0}$
are studied by employing the perturbative QCD factorization approach. We find the following results: (a) For the decays $B^-\to  a^{-}_0(1450)K^{*0}, \bar B^0\to  a^{+}_0(1450)K^{*-}, a^{0}_0(1450)\bar K^{*0}$, their branching ratios in scenario II are
larger than those in scenario I about one order. So it is easy for the experiments to differentiate between the scenario I and II for the meson
$a_0(1450)$. (b)For the decay $B^-\to  a^{0}_0(1450)K^{*-}$, due to not receiving the enhancement from the $K^*-$emission factorizable diagrams, its penguin operator
contributions are the smallest in scenario II, which makes its branching ratio drop into the order of $10^{-6}$. Even so, its branching ratio in scenario
II is still larger than that in scenario I about $2.5$ times. (c) Even though our predictions are much larger than those from the QCD factorization
results, they are still consistent with each other within the large theoretical errors from the annihilation diagrams. (d) We predict the direct CP-
violating asymmetry of the decay $B^-\to a^{-}_0(1450)K^{*0}$ is small and only a few percent.
\end{abstract}

\vspace{1cm}

\maketitle


\section{Introduction}\label{intro}
Along with many scalar mesons found in experiments, more and more efforts have been made to study the scalar meson spectrum
theoretically \cite{nato,jaffe,jwei,baru,celenza,stro,close1}. Unlike the pseudoscalar, vector, axial, and tensor mesons constant of
light quarks, which are reasonable in terms of their $SU(3)$ classification and quark content, the scalar mesons are too many to
accommodate them in one nonet. In fact, the number of the current experimentally known scalar mesons is more than 2 times that
of a nonet.  So it is believed that there are at least two nonets below and above 1 GeV. Today, it is still a difficult but interesting topic. Our most important
task is to uncover the
mysterious structure of the scalar mesons.  There are two typical schemes for the classification to
them \cite{nato,jaffe}. Scenario I (SI): the nonet mesons below
1 GeV, including $f_0(600), f_0(980), K^*(800)$ and $a_0(980)$, are
usually viewed as the lowest lying $q\bar q$ states, while the nonet
ones near 1.5 GeV, including $f_0(1370), f_0(1500)/f_0(1710),
K^*(1430)$, and $a_0(1450)$, are suggested as the first excited
states. In scenario II (SII), the nonet mesons near 1.5 GeV are
treated as $q\bar q$ ground states, while the nonet mesons below 1
GeV are exotic states beyond the quark model, such as four-quark
bound states. It should be four scalar mesons in each nonet, but there are five nonet mesons near 1.5 GeV. People generally believe that
$K^*_0(1430), a_0(1450)$ and two isosinglet scalar mesons compose one nonet, it means that one of the three isosinglet scalars $f_0(1370), f_0(1500),
f_0(1710)$ can not be explained as $q\bar q$ state and might be a scalar glueball. There are many discussions \cite{amsler,close2,hexg,ccl}
which one is most possible a scalar glueball based on a flavor-mixing scheme for these three scalar mesons, which induces there are more ambiguous about
their inner structures. By contrast, the scalar mesons $K^*_0(1430), a_0(1450)$ have been confirmed to a conventional $q\bar q$ meson in many approaches
\cite{yao,mathur,lee,bardeen}. So the calculations for the $B$ decays involved in either of these two scalar mesons in the final states should be more trustworthy.

The production of the scalar mesons from B-meson decays provides a different unique insight to the inner structures of these mesons.
It provides various factorization approaches a new usefulness.
Here we would like to use the perturbative QCD (PQCD) approach to study $a_0(1450)$
in decays $B^-\to  a^{0}_0(1450)K^{*-}, a^{-}_0(1450)\bar K^{*0}$ and $\bar B^0\to  a^{+}_0(1450)K^{*-}, a^{0}_0(1450)\bar K^{*0}$. Certainly, these decays have been
studied within the QCD
factorization approach \cite{ccysv}, in which the factorizable annihilation diagrams are calculated through a phenomenological parameter.
So there are large theoretical errors for the QCD factorization predictions. To make precise predictions of their branching ratios and
CP-violating asymmetries, it is necessary to make reliable calculations for the contributions from the factorizable annihilation diagrams. By contrast, these
diagrams are calculable within the PQCD approach effectively.

In the following, $a_0(1450)$ is denoted as $a_0$ in some places for convenience.
The layout of this paper is as follows. In Sec. \ref{proper}, the relevant decay constants
and light-cone distribution amplitudes of relevant mesons are introduced.
In Sec. \ref{results}, we then analyze these decay channels using the PQCD approach.
The numerical results and the discussions are given
in Sec. \ref{numer}. The conclusions are presented in the final part.


\section{decay constants and distribution amplitudes }\label{proper}

For the wave function of the heavy B meson,
we take
\be
\Phi_B(x,b)=
\frac{1}{\sqrt{2N_c}} (\psl_B +m_B) \gamma_5 \phi_B (x,b).
\label{bmeson}
\en
Here only the contribution of Lorentz structure $\phi_B (x,b)$ is taken into account, since the contribution
of the second Lorentz structure $\bar \phi_B$ is numerically small \cite{cdlu} and has been neglected. For the
distribution amplitude $\phi_B(x,b)$ in Eq.(\ref{bmeson}), we adopt the following model:
\be
\phi_B(x,b)=N_Bx^2(1-x)^2\exp[-\frac{M^2_Bx^2}{2\omega^2_b}-\frac{1}{2}(\omega_bb)^2],
\en
where $\omega_b$ is a free parameter, we take $\omega_b=0.4\pm0.04$ Gev in numerical calculations, and $N_B=91.745$
is the normalization factor for $\omega_b=0.4$.

In the two-quark picture, the vector decay constant $f_{a_0}$ and the
scalar decay constant $\bar {f}_{a_0}$ for the scalar meson $a_0$
can  be defined as \be \langle a_0(p)|\bar q_2\gamma_\mu
q_1|0\ra&=&f_{a_0}p_\mu, \en \be \langle a_0(p)|\bar
q_2q_1|0\ra=m_{a_0}\bar {f}_{a_0}, \label{fbar} \en where
$m_{a_0}(p)$ is the mass (momentum) of the scalar meson $a_0(1450)$. The
relation between $f_{a_0}$ and $\bar f_{a_0}$ is \be
\frac{m_{{a_0}}}{m_2(\mu)-m_1(\mu)}f_{{a_0}}=\bar f_{{a_0}}, \en
where $m_{1,2}$ are the running current quark masses. For the scalar
meson $a_0(1450)$, $f_{a_0}$ will get a very small value after the
$SU(3)$ symmetry breaking is considered. The light-cone
distribution amplitudes for the  scalar meson $a_0(1450)$
can be written as \be \langle a_0(p)|\bar q_1(z)_l
q_2(0)_j|0\rangle &=&\frac{1}{\sqrt{2N_c}}\int^1_0dx \; e^{ixp\cdot
z}\non && \times \{ p\sl\Phi_{a_0}(x)
+m_{a_0}\Phi^S_{a_0}(x)+m_{a_0}(n\sl_+n\sl_--1)\Phi^{T}_{a_0}(x)\}_{jl}.\quad\quad\label{LCDA}
\en Here $n_+$ and $n_-$ are lightlike vectors:
$n_+=(1,0,0_T),n_-=(0,1,0_T)$, and $n_+$ is parallel with the moving
direction of the scalar meson. The normalization can be related to
the decay constants: \be \int^1_0 dx\Phi_{a_0}(x)=\int^1_0
dx\Phi^{T}_{a_0}(x)=0,\,\,\,\,\,\,\,\int^1_0
dx\Phi^{S}_{a_0}(x)=\frac{\bar f_{a_0}}{2\sqrt{2N_c}}\;. \en The
twist-2 light-cone distribution amplitude $\Phi_{a_0}$ can be expanded in the Gegenbauer
polynomials: \be \Phi_{a_0}(x,\mu)&=&\frac{\bar
f_{a_0}(\mu)}{2\sqrt{2N_c}}6x(1-x)\left[B_0(\mu)+\sum_{m=1}^\infty
B_m(\mu)C^{3/2}_m(2x-1)\right], \en where the decay constants and
the Gegenbauer moments $B_1,B_3$ of distribution amplitudes for
$a_0(1450)$ have been calculated in the QCD sum rules \cite{ccysp}.
These values are all scale dependent and specified below: \be
{\rm scenario I:} B_1&=&0.89\pm0.20, B_3=-1.38\pm0.18, \bar f_{a_0}=-(280\pm30){\rm MeV},\\
{\rm scenario II:}B_1&=&-0.58\pm0.12, B_3=-0.49\pm0.15, \bar
f_{a_0}=(460\pm50){\rm MeV},\quad \en which are taken by fixing the
scale at 1GeV.

As for the twist-3 distribution amplitudes $\Phi_{a_0}^S$ and
$\Phi_{a_0}^T$, we adopt the asymptotic form: \be \Phi^S_{a_0}&=&
\frac{1}{2\sqrt {2N_c}}\bar f_{a_0},\,\,\,\,\,\,\,\Phi_{a_0}^T=
\frac{1}{2\sqrt {2N_c}}\bar f_{a_0}(1-2x). \en

For our considered
decays, the vector meson $K^*$ is longitudinally polarized. The
longitudinal polarized component of the wave function is given as
\be
\Phi_{K^*}=\frac{1}{\sqrt{2N_c}}\left\{\esl\left[m_{K^*}\Phi_{K^*}(x)+\pvsl\Phi_{K^*}^t(x)\right]+m_{K^*}\Phi^s_{K^*}(x)\right\},
\en
where the first term is the leading twist wave function (twist-2),
while the second and third term are subleading twist (twist-3) wave
functions. They can be parameterized as
\be
 \Phi_{K^*}(x) &=&  \frac{f_{K^*}}{2\sqrt{2N_c} }
    6x (1-x)
    \left[1+a_{1K^*}C^{3/2}_1(2x-1)+a_{2K^*}C^{3/2}_2(2x-1)\right],\label{piw1}
\en
\be
 \Phi^t_{K^*}(x) =   \frac{3f^T_{K^*}}{2\sqrt{2N_c}}(1-2x),\quad \Phi^s_{K^*}(x)
 =\frac{3f^T_{K^*}}{2\sqrt{2N_c}}(2x-1)^2,\label{piw}
\en
where the longitudinal decay constant $f_{K^*}=(217\pm5)$Mev and the transverse decay constant $f^T_{K^*}=(185\pm10)$Mev,  the Gegenbauer moments $a_{1K^*}=0.03, a_{2K^*}=0.11$ \cite{pball} and the Gegenbauer polynomials $C^{\nu}_n(t)$ are given as
\be
C^{3/2}_1(t)&=&3t, \qquad C^{3/2}_2(t)=\frac{3}{2}(5t^2-1).\label{eq:c124}
\en


\section{ the perturbative QCD  calculation} \label{results}

Under the two-quark model for the scalar meson $a_0(1450)$ supposition,
the decay amplitude for $B\to a_0K^*$
 can be conceptually written as the convolution,
\be
{\cal A}(B \to K^*a_0)\sim \int\!\! d^4k_1
d^4k_2 d^4k_3\ \mathrm{Tr} \left [ C(t) \Phi_B(k_1) \Phi_{K^*}(k_2)
\Phi_{a_0}(k_3) H(k_1,k_2,k_3, t) \right ], \label{eq:con1}
\en
where $k_i$'s are momenta of the antiquarks included in each meson, and
$\mathrm{Tr}$ denotes the trace over Dirac and color indices. $C(t)$
is the Wilson coefficient which results from the radiative
corrections at a short distance. In the above convolution, $C(t)$
includes the harder dynamics at a larger scale than the $M_B$ scale and
describes the evolution of local $4$-Fermi operators from $m_W$ (the
$W$ boson mass) down to the $t\sim\mathcal{O}(\sqrt{\bar{\Lambda} M_B})$
scale, where $\bar{\Lambda}\equiv M_B -m_b$. The function
$H(k_1,k_2,k_3,t)$ describes the four-quark operator and the
spectator quark connected by
 a hard gluon, whose $q^2$ is in the order
of $\bar{\Lambda} M_B$ and includes the
$\mathcal{O}(\sqrt{\bar{\Lambda} M_B})$ hard dynamics. Therefore,
this hard part $H$ can be perturbatively calculated. The function
$\Phi_{(B, K^*, a_0)}$ are the wave functions of the vector mesons $B, K^*$ and
the scalar meson $a_0$, respectively.

Since the $b$ quark is rather heavy, we consider the $B$ meson at rest
for simplicity. It is convenient to use the light-cone coordinate $(p^+,
p^-, {\bf p}_T)$ to describe the meson's momenta, \be p^\pm =
\frac{1}{\sqrt{2}} (p^0 \pm p^3), \quad {\rm and} \quad {\bf p}_T =
(p^1, p^2). \en Using these coordinates, the $B$ meson and the two
final state meson momenta can be written as \be P_B =
\frac{M_B}{\sqrt{2}} (1,1,{\bf 0}_T), \quad P_{2} =
\frac{M_B}{\sqrt{2}}(1-r^2_{a_0},r^2_{K^*},{\bf 0}_T), \quad P_{3} =
\frac{M_B}{\sqrt{2}} (r^2_{a_0},1-r^2_{K^*},{\bf 0}_T), \en respectively, where the ratio $r_{a_0(K^*)}=m_{a_0(K^*)}/M_B$, and
$m_{a_0(K^*)}$ is the scalar meson $a_0$ (the vector meson $K^*$) mass. Putting the antiquark momenta in $B$,
$K^*$, and $a_0$ mesons as $k_1$, $k_2$, and $k_3$, respectively, we can
choose
\be k_1 = (x_1 P_1^+,0,{\bf k}_{1T}), \quad k_2 = (x_2
P_2^+,0,{\bf k}_{2T}), \quad k_3 = (0, x_3 P_3^-,{\bf k}_{3T}). \en
For these considered decay channels, the integration over $k_1^-$,
$k_2^-$, and $k_3^+$ in Eq.(\ref{eq:con1}) will lead to
\be
 {\cal
A}(B \to K^*a_0) &\sim &\int\!\! d x_1 d x_2 d x_3 b_1 d b_1 b_2 d
b_2 b_3 d b_3 \non && \cdot \mathrm{Tr} \left [ C(t) \Phi_B(x_1,b_1)
\Phi_{K^*}(x_2,b_2) \Phi_{a_0}(x_3, b_3) H(x_i, b_i, t) S_t(x_i)\,
e^{-S(t)} \right ], \quad \label{eq:a2}
\en
where $b_i$ is the
conjugate space coordinate of $k_{iT}$, and $t$ is the largest
energy scale in function $H(x_i,b_i,t)$.
In order to smear the end-point singularity on $x_i$,
the jet function $S_t(x)$ \cite{li02}, which comes from the
resummation of the double logarithms $\ln^2x_i$, is used.
The last term $e^{-S(t)}$ in Eq.(\ref{eq:a2}) is the Sudakov form factor which suppresses
the soft dynamics effectively \cite{soft}.

 For the considered decays, the related weak effective
Hamiltonian $H_{eff}$ can be written as \cite{buras96}
\be
\label{eq:heff} {\cal H}_{eff} = \frac{G_{F}} {\sqrt{2}} \,
\left[\sum_{p=u,c}V_{pb} V_{ps}^* \left (C_1(\mu) O_1^p(\mu) +
C_2(\mu) O_2^p(\mu) \right) -V_{tb} V_{ts}^*\sum_{i=3}^{10} C_{i}(\mu) \,O_i(\mu)
\right] \;,
\en
where the Fermi constant $G_{F}=1.166 39\times
10^{-5} GeV^{-2}$ and the functions $Q_i (i=1,...,10)$ are the local four-quark operators. We specify below
the operators in ${\cal H}_{eff}$ for $b \to s$ transition: \be
\begin{array}{llllll}
O_1^{u} & = & \bar s_\alpha\gamma^\mu L u_\beta\cdot \bar
u_\beta\gamma_\mu L b_\alpha\ , &O_2^{u} & = &\bar
s_\alpha\gamma^\mu L u_\alpha\cdot \bar
u_\beta\gamma_\mu L b_\beta\ , \\
O_3 & = & \bar s_\alpha\gamma^\mu L b_\alpha\cdot \sum_{q'}\bar
 q_\beta'\gamma_\mu L q_\beta'\ ,   &
O_4 & = & \bar s_\alpha\gamma^\mu L b_\beta\cdot \sum_{q'}\bar
q_\beta'\gamma_\mu L q_\alpha'\ , \\
O_5 & = & \bar s_\alpha\gamma^\mu L b_\alpha\cdot \sum_{q'}\bar
q_\beta'\gamma_\mu R q_\beta'\ ,   & O_6 & = & \bar
s_\alpha\gamma^\mu L b_\beta\cdot \sum_{q'}\bar
q_\beta'\gamma_\mu R q_\alpha'\ , \\
O_7 & = & \frac{3}{2}\bar s_\alpha\gamma^\mu L b_\alpha\cdot
\sum_{q'}e_{q'}\bar q_\beta'\gamma_\mu R q_\beta'\ ,   & O_8 & = &
\frac{3}{2}\bar s_\alpha\gamma^\mu L b_\beta\cdot
\sum_{q'}e_{q'}\bar q_\beta'\gamma_\mu R q_\alpha'\ , \\
O_9 & = & \frac{3}{2}\bar s_\alpha\gamma^\mu L b_\alpha\cdot
\sum_{q'}e_{q'}\bar q_\beta'\gamma_\mu L q_\beta'\ ,   & O_{10} & =
& \frac{3}{2}\bar s_\alpha\gamma^\mu L b_\beta\cdot
\sum_{q'}e_{q'}\bar q_\beta'\gamma_\mu L q_\alpha'\ ,
\label{eq:operators} \end{array}
\en
where $\alpha$ and $\beta$ are
the $SU(3)$ color indices; $L$ and $R$ are the left- and
right-handed projection operators with $L=(1 - \gamma_5)$, $R= (1 +
\gamma_5)$. The sum over $q'$ runs over the quark fields that are
active at the scale $\mu=O(m_b)$, i.e., $(q'\epsilon\{u,d,s,c,b\})$.


\begin{figure}[t,b]
\vspace{-3cm} \centerline{\epsfxsize=16 cm \epsffile{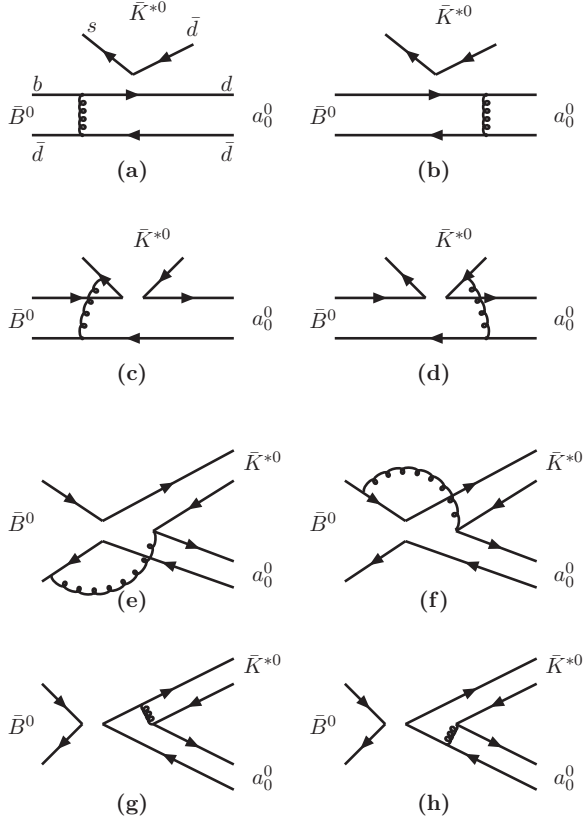}}
\vspace{-9cm} \caption{ Diagrams contributing to the decay $\bar{B}^0\to \bar K^{*0}
a^0_0(1450)$ .}
 \label{fig1}
\end{figure}

In Fig.~1, we give the leading order Feynman diagrams for the channel $\bar{B}^0\to a^0_0(1450)K^{*0}$ as an example. For the
fractorizable and nonfactorizable emission diagrams Fig.1(a), 1(b) and 1(c), 1(d), if one exchanges the $K^{*0}$ and $a^0_0$, the
corresponding diagrams also exist. But there are not this kind of exchange diagrams for the factorizable and nonfactorizable annihilation
diagrams, that is
Fig.1 (e), 1(f) and 1(g), 1(h). If we replace the $\bar d$ quark in both $\bar B^0$ and $a^0_0$ with $\bar u$ quark, we will get
the Feynman diagrams for the decay $B^-\to a^-_0(1450)K^{*0}$. If we replace the $d(\bar d)$ quark in $a_0^0 (K^{*0})$ with $u (\bar u)$, we will get
the Feynman diagrams for the decay $\bar B^0\to a^+_0(1450)K^{*-}$. While there are not the diagrams obtained by exchanging
the two final state mesons for these two channels. For the decay $B^-\to a^0_0(1450)K^{*-}$, its Feynman diagrams are distinctive: the meson $a_0(1450)$
is emitted (the upper meson) in the fractorizable (nonfactorizable) emission diagrams, while the meson $K^*$ is the upper meson in the
fractorizable (nonfactorizable)
annihilation diagrams.
The detailed analytic
formulae for the diagrams of each decay are not presented and can be obtained from those of $B\to f_0(980)K^*$ \cite{zqzhang} by
replacing corresponding wave functions and parameters.

Combining the contributions from different diagrams, the total decay
amplitudes for these decays can be written as
\be
\sqrt{2}{\cal M}(\bar K^{*0}a^0_0)&=&\xi_uM_{eK^*}C_2-\xi_t\left[M_{eK^*}\frac{3C_{10}}{2}
+M^{P2}_{eK^*}\frac{3C_{8}}{2}-(F_{ea_0}+F_{aa_0})\left(a_4-\frac{a_{10}}{2}\right)
\right.\non &&\left.-(M_{ea_0}+M_{aa_0})\left(C_3-\frac{1}{2}C_9\right)
-(M^{P1}_{ea_0}+M^{P1}_{aa_0})\left(C_5-\frac{1}{2}C_7\right)\right.\non &&\left.-F^{P2}_{aa_0}(a_6-\frac{1}{2}a_8)\right],\\
{\cal M}(\bar K^{*0}a^-_0)&=&\xi_u\left[M_{aa_0}C_1+F_{aa_0}a_1\right]-\xi_t\left[F_{ea_0}
\left(a_4-\frac{a_{10}}{2}\right)+F_{aa_0}\left(a_4+a_{10}\right)
\right.\non &&\left.+M_{ea_0}\left(C_3-\frac{1}{2}C_9\right)+M_{aa_0}\left(C_3+C_9\right)
+M^{P1}_{ea_0}\left(C_5-\frac{1}{2}C_7\right)
\right.\non &&\left.+M^{P1}_{aa_0}\left(C_5+C_7\right)+F^{P2}_{aa_0}(a_6+a_8)\right],\\
\sqrt{2}{\cal M}(\bar K^{*-}a^0_0)&=&\xi_u\left[M_{eK^*}C_2+M_{aa_0}C_1+F_{aa_0}a_1\right]-\xi_t\left[
M_{eK^*}\frac{3}{2}C_{10}+M^{P2}_{eK^*}\frac{3}{2}C_{8}\right.\non &&\left.+M_{aa_0}\left(C_3+C_9\right)+
M^{P1}_{aa_0}\left(C_5+C_7\right)\right.\non &&\left.+F_{aa_0}\left(a_4+a_{10}\right)
+F^{P2}_{aa_0}(a_6+a_8)\right],\\
{\cal M}(\bar K^{*-}a^+_0)&=&\xi_u\left[F_{ea_0}a_1+M_{ea_0}C_1\right]-\xi_t\left[F_{ea_0}\left(a_4+a_{10}\right)+
M_{ea_0}\left(C_3+C_9\right)\right.\non &&\left.+M^{P1}_{ea_0}\left(C_5+C_7\right)+M_{aa_0}\left(C_3-\frac{1}{2}C_9\right)+
M^{P1}_{aa_0}\left(C_5-\frac{1}{2}C_7\right)\right.\non &&\left.+F_{aa_0}\left(a_4-\frac{1}{2}a_{10}\right)
+F^{P2}_{aa_0}\left(a_6-\frac{1}{2}a_8\right)\right],
\en
The
combinations of the Wilson coefficients are defined as usual
\cite{zjxiao}:
 \be
a_{1}(\mu)&=&C_2(\mu)+\frac{C_1(\mu)}{3}, \quad
a_2(\mu)=C_1(\mu)+\frac{C_2(\mu)}{3},\non
a_i(\mu)&=&C_i(\mu)+\frac{C_{i+1}(\mu)}{3},\quad
i=3,5,7,9,\non
a_i(\mu)&=&C_i(\mu)+\frac{C_{i-1}(\mu)}{3},\quad
i=4, 6, 8, 10.\label{eq:aai} \en

\section{Numerical results and discussions} \label{numer}

We use the following input parameters in the numerical calculations \cite{pdg08,barbar}:
\be
f_B&=&190 MeV, M_B=5.28 GeV, M_W=80.41 GeV,\\
V_{ub}&=&|V_{ub}|e^{-i\gamma}=3.93\times10^{-3}e^{-i68^\circ},\\
V_{us}&=&0.2255, V_{tb}=1.0, V_{ts}=0.0387, \\
\tau_{B^\pm}&=&1.638\times 10^{-12} s,\tau_{B^0}=1.530\times 10^{-12} s.
\en
Using the wave functions and the values of relevant input parameters, we find the numerical values
of the form factor $B\to a_0(1450)$ at zero momentum transfer:
\be
F^{\bar B^0\to a_0}_0(q^2=0)&=&-0.42^{+0.04+0.04+0.05+0.06}_{-0.03-0.03-0.04-0.07}, \quad\mbox{ scenario I},
\\F^{\bar B^0\to a_0}_0(q^2=0)&=&0.86^{+0.04+0.05+0.10+0.14}_{-0.03-0.04-0.09-0.11}, \quad\;\;\;\mbox{ scenario II},
\en
where the uncertainties are mainly from the Gegenbauer moments $B_1$, $B_3$, the decay constant of the meson $a_0(1450)$, the $B$-meson shape parameter
$\omega=0.40\pm0.04$ GeV. These predictions are larger than those given in Ref.\cite{lirh}, for using different values for the threshold parameter $c$
in the jet function. Certainly, they are consistent with each other in errors.

\begin{figure}[t,b]
\begin{center}
\includegraphics[scale=0.7]{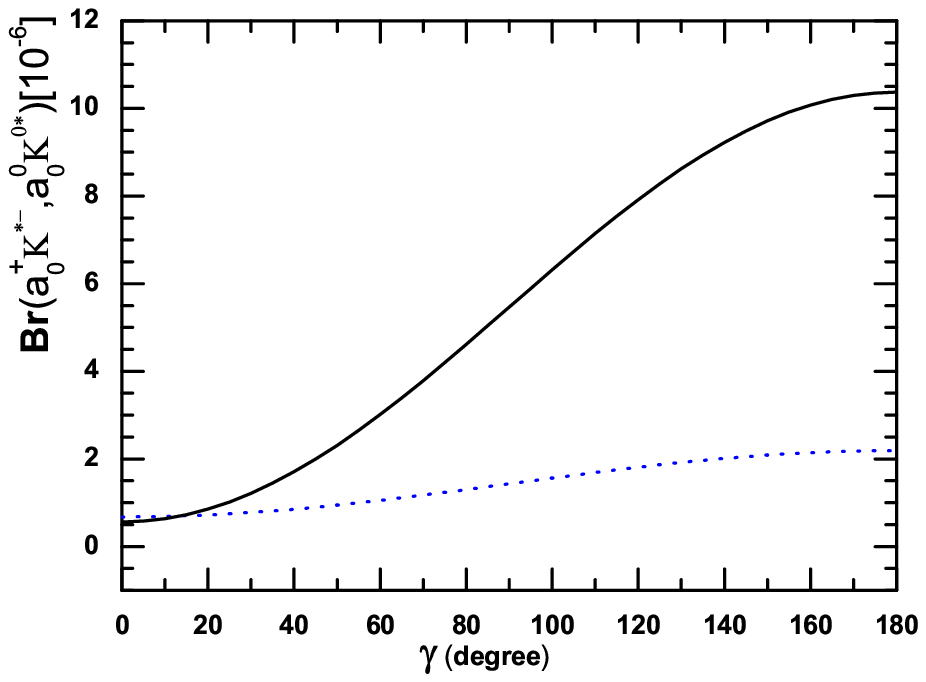}
\includegraphics[scale=0.7]{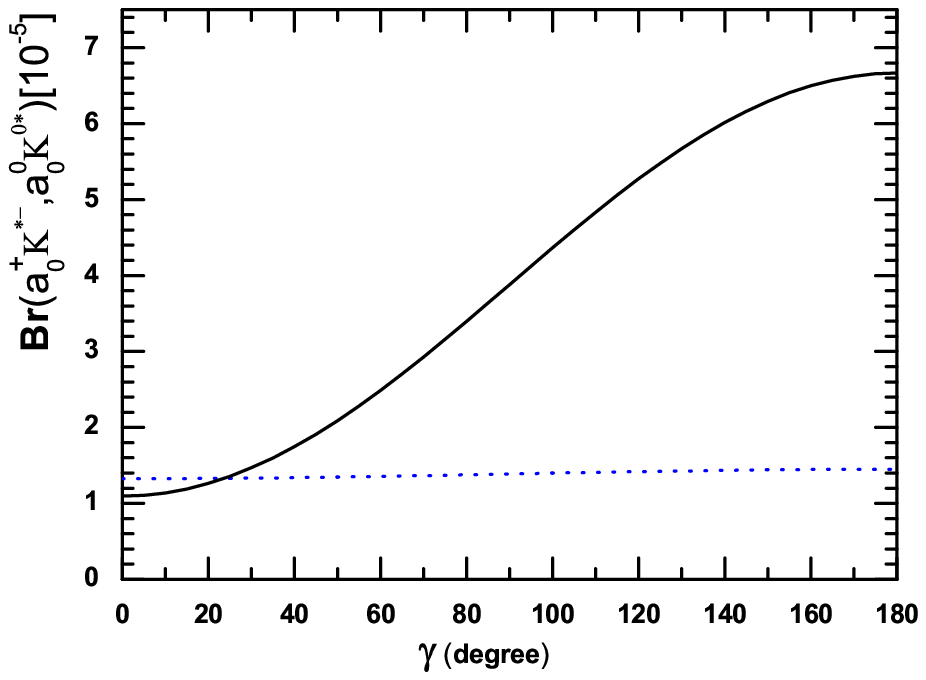}
\vspace{0.3cm} \caption{The dependence of the branching ratios for
$\bar B^0\to  a^{+}_0(1450)K^{*-}$ (solid curve) and $\bar B^0\to  a^{0}_0(1450)\bar K^{*0}$ (dotted curve) on the
Cabibbo-Kobayashi-Maskawa angle $\gamma$. The left (right) panel is plotted in scenario I (II).}\label{fig2}
\end{center}
\end{figure}

\begin{figure}[t,b]
\begin{center}
\includegraphics[scale=0.7]{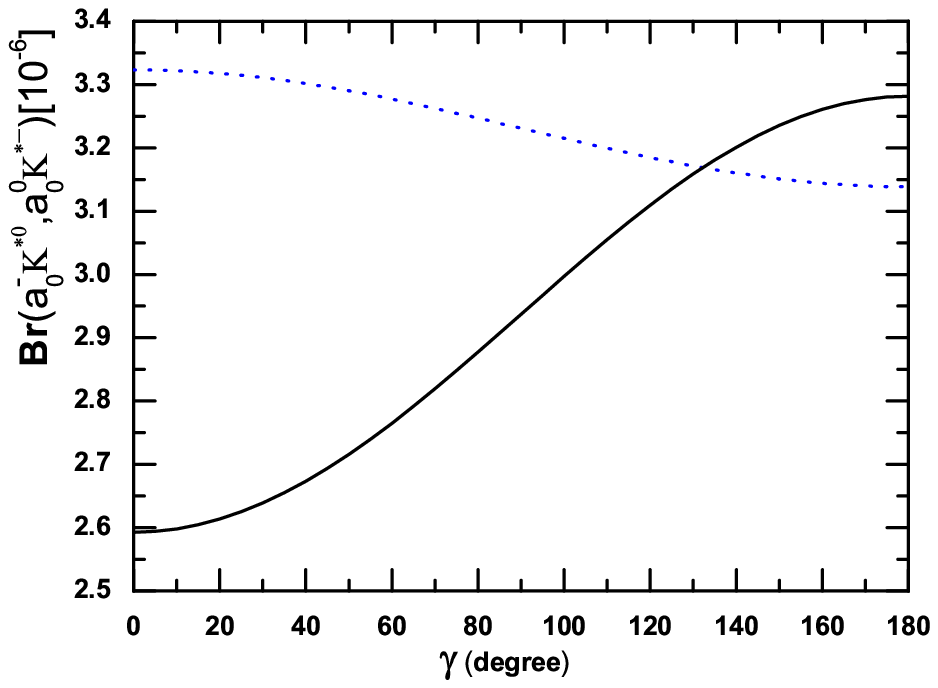}
\includegraphics[scale=0.7]{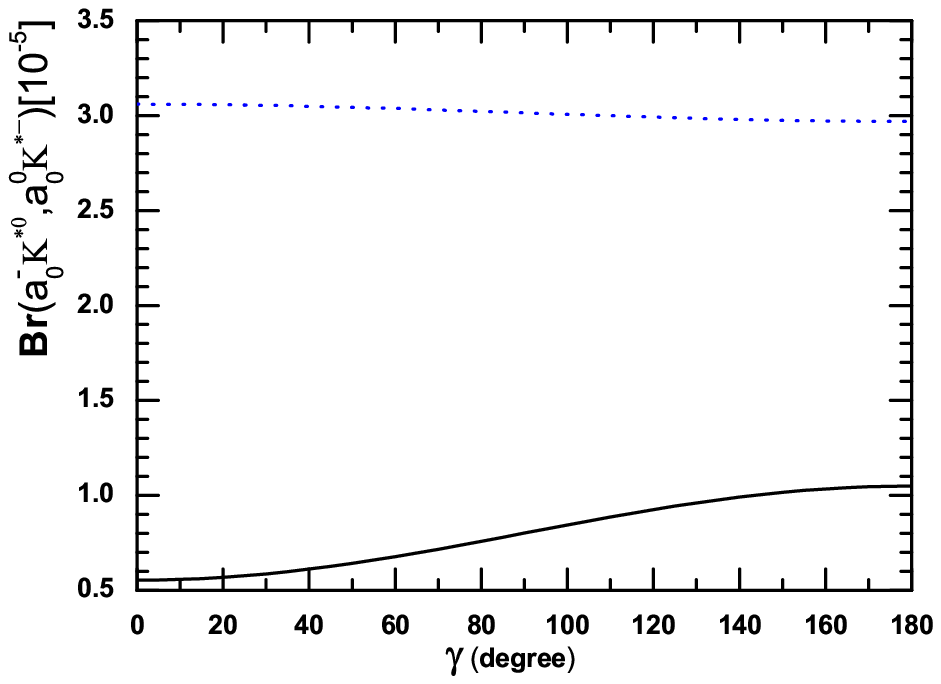}
\vspace{0.3cm} \caption{The dependence of the branching ratios for
$B^-\to  a^{0}_0(1450)K^{*-}$ (solid curve) and $B^-\to  a^{-}_0(1450)\bar K^{*0}$ (dotted curve) on the
Cabibbo-Kobayashi-Maskawa angle $\gamma$. The left (right) panel is plotted in scenario I (II).}\label{fig3}
\end{center}
\end{figure}

In the B-rest frame, the decay rates of $B\to a_0(1450)K^*$ can be written as
\be
\Gamma=\frac{G_F^2}{32\pi m_B}|{\cal M}|^2(1-r^2_{a_0}),
\en
where ${\cal M}$ is the total decay amplitude of each
considered decay and $r_{a_0}$ the mass ratio, which have been given  in  Sec. \ref{results}.
${\cal M}$ can be rewritten as
\be
{\cal M}= V_{ub}V^*_{us}T-V_{tb}V^*_{ts}P=V_{ub}V^*_{us}\left[1+ze^{i(\delta-\gamma)}\right] \label{ampde},
\en
where $\gamma$ is the Cabibbo-Kobayashi-Maskawa weak phase angle, and $\delta$ is the relative strong phase between
the tree and the penguin amplitudes, which are denote as "T" and "P", respectively. The term $z$ describes the ratio of penguin to tree
contributions and is defined as
\be
z=\left|\frac{V_{tb}V^*_{ts}}{V_{ub}V^*_{us}}\right|\left|\frac{P}{T}\right|.
\en
From Eq.(\ref{ampde}), it is easy to write decay amplitude $\overline {\cal M}$ for the corresponding conjugated decay mode. So the CP-averaged
branching ratio for each considered decay is defined as
\be
{\cal B}=(|{\cal M}|^2+|\overline{\cal M}|^2)/2=|V_{ub}V^*_{us}T|^2\left[1+2z\cos\gamma\cos\delta+z^2\right].\label{brann}
\en
Using the input parameters and the wave functions as specified in this and previous sections, it is easy to get the branching ratios in two scenarios:
\be
{\cal B}(B^-\to  a^{0}_0(1450)K^{*-})=(2.8^{+0.4+1.0+0.6+0.1}_{-0.4-0.0-0.6-0.1})\times 10^{-6}, Scenario I,\\
{\cal B}(B^-\to  a^{-}_0(1450)\bar K^{*0})=(3.3^{+0.6+0.4+0.8+2.7}_{-0.4-0.3-0.7-1.5})\times 10^{-6}, Scenario I,\\
{\cal B}(\bar B^0\to  a^{+}_0(1450)K^{*-})=(3.6^{+0.6+0.3+0.8+2.0}_{-0.6-0.1-0.7-1.1})\times 10^{-6}, Scenario I,\\
{\cal B}(\bar B^0\to  a^{0}_0(1450)\bar K^{*0})=(1.2^{+0.1+0.1+0.2+1.0}_{-0.1-0.2-0.3-0.6})\times 10^{-6}, Scenario I;\\
{\cal B}(B^-\to  a^{0}_0(1450)K^{*-})=(7.0^{+0.9+1.6+1.7+0.2}_{-0.7-1.1-1.4-0.0})\times 10^{-6}, Scenario II,\\
{\cal B}(B^-\to  a^{-}_0(1450)\bar K^{*0})=(3.0^{+0.2+0.2+0.7+1.2}_{-0.1-0.1-0.6-0.7})\times 10^{-5}, Scenario II,\\
{\cal B}(\bar B^0\to  a^{+}_0(1450)K^{*-})=(2.8^{+0.3+0.1+0.7+0.8}_{-0.3-0.0-0.5-0.6})\times 10^{-5}, Scenario II,\\
{\cal B}(\bar B^0\to  a^{0}_0(1450)\bar K^{*0})=(1.4^{+0.1+0.0+0.3+0.5}_{-0.1-0.1-0.3-0.4})\times 10^{-5}, Scenario II.
\en
In the above results, the first two errors come from the uncertainties of
the Gegenbauer moments $B_1$, $B_3$ of the scalar meson, and the third one is from the decay constant of $a_0(1450)$. The
last one comes from the uncertainty in the $B$ meson shape parameter $\omega_b=0.40\pm0.04$ GeV. We also show the dependence of
the branching ratios for these considered decays on the Cabibbo-Kobayashi-Maskawa angle $\gamma$ in Fig. \ref{fig2}
and Fig. \ref{fig3}.

The branching ratios predicted by QCD factorization approach for these considered decays in scenario II are listed as \cite{ccysv}
\be
{\cal B}(B^-\to  a^{0}_0(1450)K^{*-})&=&(2.2^{+4.9+0.7+22.5}_{-4.0-0.6-8.3})\times 10^{-6}, \label{qcdf1}\\
{\cal B}(B^-\to  a^{-}_0(1450)\bar K^{*0})&=&(7.8^{+14.3+0.9+23.4}_{-11.0-0.7-9.1})\times 10^{-6}, \\
{\cal B}(\bar B^0\to  a^{+}_0(1450)K^{*-})&=&(4.7^{+4.4+1.0+14.6}_{-3.7-0.8-5.3})\times 10^{-6}, \\
{\cal B}(\bar B^0\to  a^{0}_0(1450)\bar K^{*0})&=&(2.5^{+4.4+1.0+14.6}_{-3.7-0.8-5.3})\times 10^{-6}. \label{qcdf4}
\en
Though it is well known that the annihilation diagram contributions to charmless hadronic B decays are power suppressed in the heavy-quark limit,
as emphasized in \cite{keum}, these contributions may be important for some B meson decays, here considered channels are just this kind of decays.
For this kind decays, the factorizable annihilation diagrams almost guide the final branching ratios, so it is important to calculate correctly
the amplitudes from these diagrams. While the annihilation amplitude has endpoint divergence even at twist-2 level in QCD factorization calculations, and
one cannot compute it in a self-consistent way and has to parameterize phenomenologically the endpoint divergence. So it is difficult to avoid
to bring many uncertainties to the final results. In fact, the major uncertainties listed in Eq.(\ref{qcdf1}-\ref{qcdf4}) are just from the contributions
of annihilation diagrams. Comparing with QCD factorization approach, PQCD approach can make a reliable calculation from factorizable annihilation diagrams
in $k_T$ factorization \cite{yu}. The endpoint singularity occurred in QCD factorization approach is cured here by the Sudakov factor.
Because of the large uncertainties from QCD factorization approach, our predictions in scenario II are also in agreement with the QCD factorization results within theoretical errors.

In Table \ref{amppp}, we list the values of the factorizable and nonfactorizable amplitudes from the emission and annihilation topology diagrams
of the considered decays in both scenarios. $F_e(a)a_0$ and $M_e(a)a_0$ are the $K^*-$meson emission (annihilation) factorizable contributions
nonfactorizable contributions from penguin operators, respectively. The upper label $T$ denotes the contributions from tree operators. For the
decays $B^-\to  a^{0}_0K^{*-}$ and $\bar B^0\to a^{0}_0\bar K^{*0}$, there also exists the contributions from $a_0$ emission nonfactorizable
diagrams.
\begin{table}
\caption{ Decay amplitudes for decays $B^-\to a^{0}_0K^{*-}, a^{-}_0\bar K^{*0}$, $\bar B^0\to a^{+}_0K^{*-}, a^{0}_0\bar K^{*0}$
($\times 10^{-2} \mbox {GeV}^3$) in the two scenarios.}
\begin{center}
\begin{tabular}{cc|c|c|c|c|c|c|c|c|c}
\hline \hline  &&$F^T_{ea_0}$ &$F_{ea_0}$ & $M^T_{ea_0}+M^T_{eK^*}$ &$M_{ea_0}+M_{eK^*}$& $M^T_{aa_0}$ & $M_{aa_0}$ &$F^T_{aa_0}$&$F_{aa_0}$\\
\hline
$ a^{0}_0K^{*-}$ (SI) &&...&...&$-32.6+59.5i$&$-0.14+0.30i$&$-1.8+3.2i$ &$0.11-0.06i$&$-0.5-3.5i$&$5.3+1.5i$\\
$  a^{-}_0\bar K^{*0}$ (SI) &&...&-12.5&$...$&$-0.27+0.00i$&$-2.5+4.5i$&$0.16-0.08i$ &$-0.7-4.9i$&$7.1+2.6i$\\
$ a^{+}_0K^{*-}$(SI) &&272.8&-12.0&$11.3-8.3i$&$0.02-0.28i$&...&$0.20-0.19i$&...&$7.3+2.4i$\\
$  a^{0}_0\bar K^{*0}$(SI) &&...&8.9&$-32.6+59.5i$&$0.07+0.31i$&... &$-0.14+0.13i$&...&$-5.2-1.9i$\\
$ a^{0}_0K^{*-}$ (SII) &&...&...&$47.9+39.0i$&$0.23+0.18i$&$6.7+1.6i$ &$-0.25-0.17i$&$0.6+1.1i$&$-5.2-7.9i$\\
$ a^{-}_0\bar K^{*0}$ (SII) &&...&22.9&...&$-0.73+0.70i$&$9.5+2.3i$&$-0.36-0.25i$ &$1.0+1.5i$&$-6.9-11.6i$\\
$ a^{+}_0K^{*-}$(SII) &&-548.5&22.1&$-1.5-6.4i$&$-0.84+0.57i$&...&$-0.56-0.30i$&...&$-7.1-11.6i$\\
$ a^{0}_0\bar K^{*0}$(SII) &&...&-16.2&$47.9+39.0i$&$0.72-0.33i$&...&$0.40+0.21i$&...&$5.0+8.2i$\\
\hline\hline
\end{tabular}\label{amppp}
\end{center}
\end{table}

In order to show the importance of the contributions from penguin operators, we can show the branching ratio in another way:
\be
{\cal B}=|V_{ub}V^*_{us}|^2(T_r^2+T_i^2)
+|V_{tb}V^*_{ts}|^2(P_r^2+P_i^2)-|V_{ub}V^*_{us}V_{tb}V^*_{ts}|\cos\gamma(T_rP_r+T_iP_i).\label{brann1}
\en
If the both sides of the upper equation are divided by the constant $|V_{ub}V^*_{us}|^2$, one can get
\be
\frac{{\cal B}}{|V_{ub}V^*_{us}|^2}&=&(T_r^2+T_i^2)
+|\frac{V_{tb}V^*_{ts}}{V_{ub}V^*_{us}}|^2(P_r^2+P_i^2)-|\frac{V_{tb}V^*_{ts}}{V_{ub}V^*_{us}}|\cos\gamma(T_rP_r+T_iP_i)\non\label{brann2}
&=&(T_r^2+T_i^2)+1936(P_r^2+P_i^2)-16.3(T_rP_r+T_iP_i).
\en
From Eq.(\ref{brann2}), we can find the contributions from tree operators are strongly CKM-suppressed compared with those from penguin
operators. Certainly, the contributions from the conference of tree and penguin operators are also small. So generally speaking, the
branching ratios are proportional to $(P_r^2+P_i^2)$, that is to say if ones penguin operator contributions are large, its branching
ratio is also large. But the branching ratio of $\bar B^0\to  a^{+}_0(1450)K^{*-}$ for scenario I is excepted. It is because the contributions from
tree operators are enhanced very much by the large Wilson coefficients $a_1=C_2+C_1/3$, which results they are very large to survive the
aforementioned suppression. So exactly speaking, the mode $\bar B^0\to  a^{+}_0(1450)K^{*-}$ is a tree-dominated decay in scenario I. On the other side,
the conferences from tree and penguin operators also strengthen the final result. So, even though
the contributions from the penguin operators for this channel are the smallest in scenario I, instead, it receives a larger branching ratio
. Another abnormal decay channel is $B^-\to  a^{0}_0(1450)K^{*-}$. In scenario II, the branching ratios of other three decays
are at the order of $10^{-5}$, while the branching ratio of decay $B^-\to  a^{0}_0(1450)K^{*-}$ is the smallest one and only a few times $10^{-6}$.
The reason is that the contributions from penguin operators of this decay are the smallest. Compared with the decay $B^-\to  a^{-}_0(1450)\bar K^{*0}$
though, the decay mode $a^{0}_0(1450)K^{*-}$ receives extra tree contributions $M^T_{eK^*}+M^T_{eK^*}$, which makes its total tree contribution
almost 7 times larger than that of
the mode $a^{-}_0(1450)\bar K^{*0}$, while as mentioned above, the tree contributions are strongly suppressed and not much helpful to
enhance the branching ratio. Compared with other three decays, the decay $B^-\to  a^{0}_0(1450)K^{*-}$ does not receive the enhancement
from the $K^{*}$-emission factorizable diagrams and get the smallest contributions from the penguin operators, which makes its branching ratio
curve shown in Fig. 3 drop a lot. Certainly, the mode $a^{0}_0(1450)K^{*-}$ does not receive this kind of enhancement (that is $F_{ea_0}$)
in scenario I, too. In fact, $F_{ea_0}$ and $F_{aa_0}$ shown in Table I are destructive for the other three decays in both scenarios.
The destruction induces the mode $a^{0}_0\bar K^{*0}$ receives a smaller penguin amplitude compared with the mode $a^{0}_0(1450)K^{*-}$ in scenario I.

Now we turn to the evaluations of the direct CP-violating asymmetries of
the considered decays in PQCD approach. The direct CP-violating asymmetry can be defined as
\be
\acp^{dir}=\frac{ |\overline{\cal M}|^2-|{\cal M}|^2 }{
 |{\cal M}|^2+|\overline{\cal M}|^2}=\frac{2z\sin\gamma\sin\delta}
{1+2z\cos\gamma\cos\delta+z^2}\;.
\en

\begin{figure}[t,b]
\begin{center}
\includegraphics[scale=0.7]{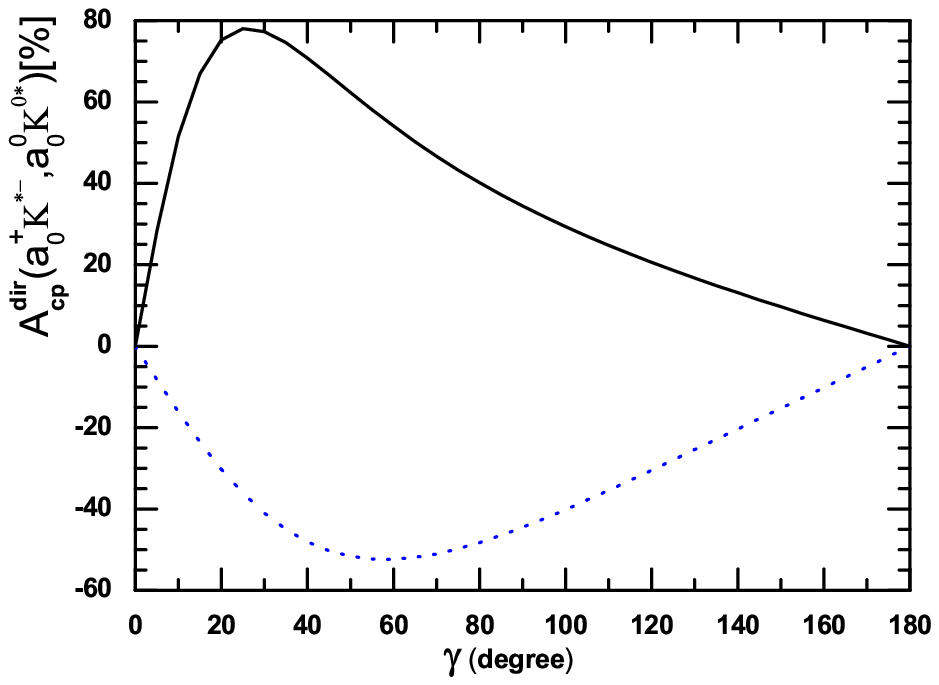}
\includegraphics[scale=0.7]{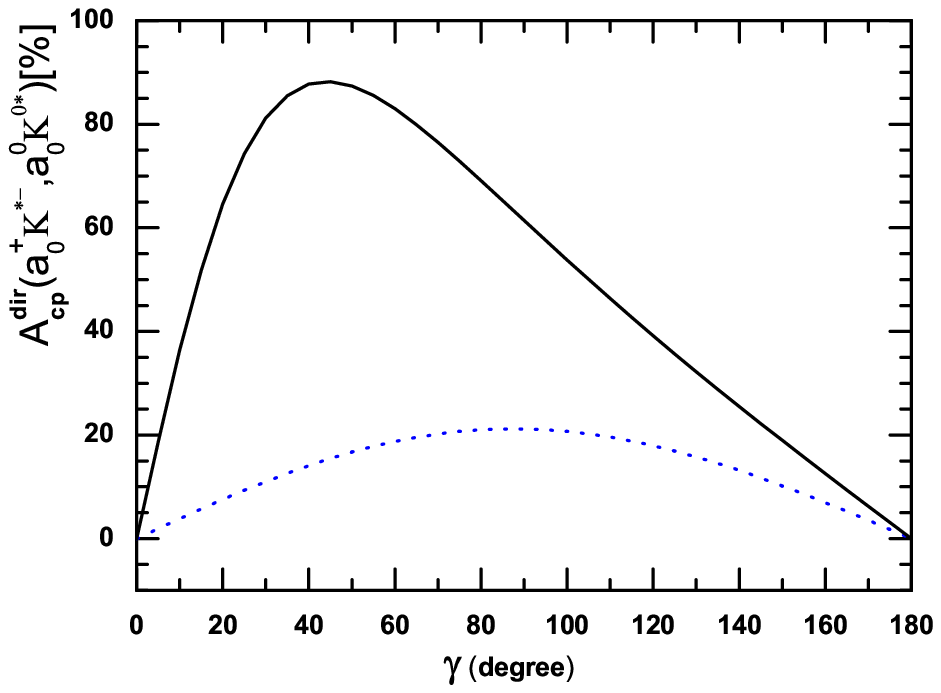}
\vspace{0.3cm} \caption{Direct CP-violating asymmetries of
$\bar B^0\to  a^{+}_0(1450)K^{*-}$ (solid curve) and $\bar B^0\to  a^{0}_0(1450)\bar K^{*0}$ (dotted curve), as functions of the
Cabibbo-Kobayashi-Maskawa angle $\gamma$. The left (right) panel is plotted in scenario I (II).}\label{fig4}
\end{center}
\end{figure}
Using the input parameters and the wave functions as specified in this and previous sections, one can find the PQCD predictions (in units of $10^{-2}$) for the direct CP-violating asymmetries of the
considered decays
\be
\acp^{dir}(B^-\to  a^{0}_0(1450)K^{*-})=-50.1^{+8.4+4.7+0.5+2.6}_{-8.8-0.0-0.0-0.6}, \mbox{ scenario I},\\
\acp^{dir}(B^-\to  a^{-}_0(1450)\bar K^{*0})=2.0^{+1.5+0.4+0.2+0.8}_{-0.8-0.0-0.0-0.1},\mbox{ scenario I},\\
\acp^{dir}(\bar B^0\to  a^{+}_0(1450)K^{*-})=48.0^{+19.2+1.5+0.1+13.5}_{-20.5-6.1-0.0-12.8},\mbox{ scenario I},\\
\acp^{dir}(\bar B^0\to  a^{0}_0(1450)\bar K^{*0})=-50.8^{+20.2+0.1+0.0+12.7}_{-20.5-0.1-0.6-11.9},\mbox{ scenario I},\\
\acp^{dir}(B^-\to  a^{0}_0(1450)K^{*-})=-11.4^{+1.5+0.3+0.1+2.0}_{-1.7-0.3-0.0-1.1}, \mbox{ scenario II},\\
\acp^{dir}(B^-\to  a^{-}_0(1450)\bar K^{*0})=-1.8^{+0.3+0.2+0.0+0.2}_{-0.2-0.1-0.0-0.2},\mbox{ scenario II},\\
\acp^{dir}(\bar B^0\to  a^{+}_0(1450)K^{*-})=78.0^{+2.2+4.8+0.0+6.6}_{-2.5-5.4-0.1-8.8},\mbox{ scenario II},\\
\acp^{dir}(\bar B^0\to  a^{0}_0(1450)\bar K^{*0})=20.0^{+2.5+1.7+0.0+0.8}_{-3.1-1.7-0.0-1.3},\mbox{ scenario II}.
\en
The main errors are induced by the uncertainties of $B_1$ and $B_3$ of $a_0(1450)$, $f_{a_0}$ and $B$ meson shape parameter $\omega_b$.

\begin{figure}[t,b]
\begin{center}
\includegraphics[scale=0.7]{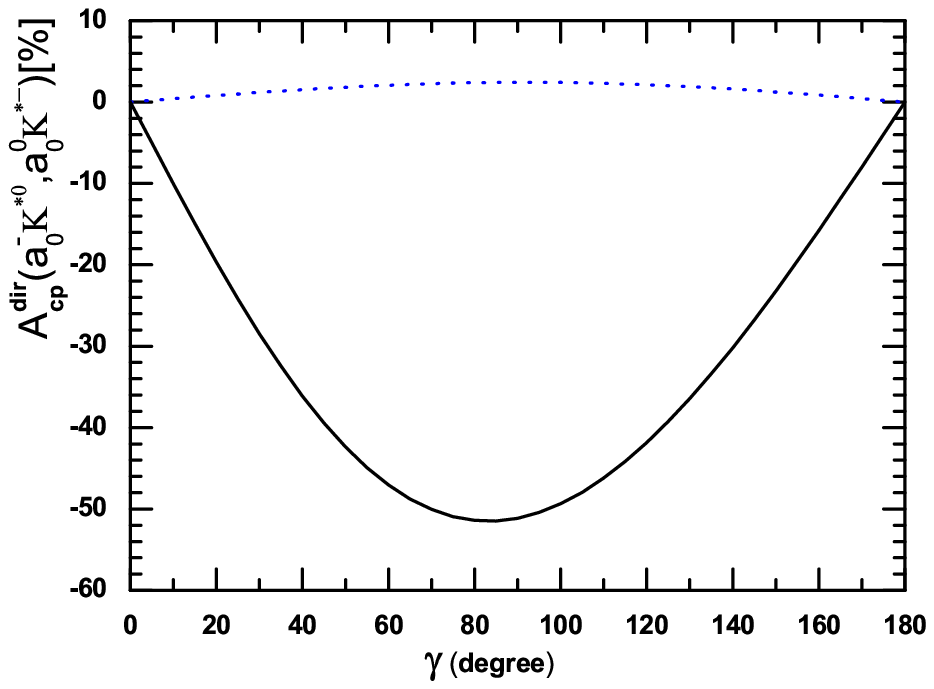}
\includegraphics[scale=0.7]{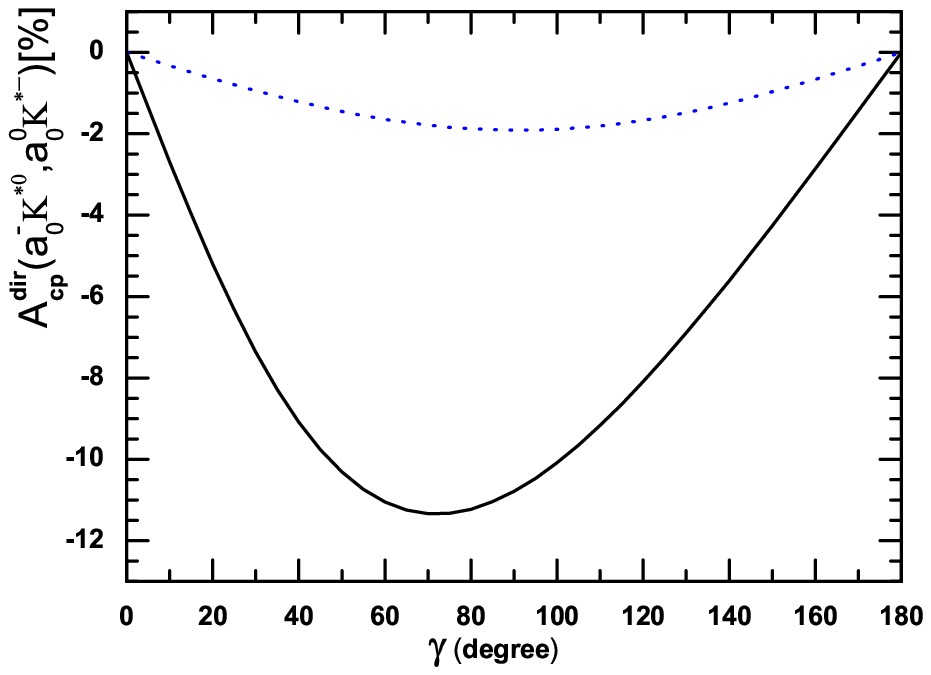}
\vspace{0.3cm} \caption{Direct CP-violating asymmetries of
$B^-\to  a^{0}_0(1450)K^{*-}$ (solid curve) and $B^-\to  a^{-}_0(1450)\bar K^{*0}$ (dotted curve) as functions of the
Cabibbo-Kobayashi-Maskawa angle $\gamma$. The left (right) panel is plotted in scenario I (II).}\label{fig5}
\end{center}
\end{figure}

The direct CP-violating asymmetries of these considered decays are displayed in Fig. \ref{fig4} and Fig. \ref{fig5}. From these figures, one can find
the direct CP-violating asymmetries of the decays $\bar B^0\to  a^{+}_0(1450)K^{*-}$ and $B^-\to  a^{0}_0(1450)K^{*-}$ have the same
sign in the two scenarios, while those of the decays $B^-\to  a^{-}_0(1450)\bar K^{*0}$ and $\bar B^0\to  a^{0}_0(1450)\bar K^{*0}$ have
contrary signs in the two scenarios. If the value of $z$ [defined in Eq.(\ref{ampde})] is very large, for example, $z_{a^{-}_0\bar K^{*0}}=91.1$ (scenario I) and $78.8$
(scenario II), the corresponding direct CP-violating asymmetry will be very small and only a few percent. If the value of $z$ is small
and only a few, for example, $z_{a^{0}_0K^{*-}}=6.2$ (scenario II) and $z_{a^{0}_0\bar K^{*0}}=9.2$ (scenario II), the corresponding
direct CP-violating asymmetry is large. If the value of $z$ is very small and not far away from 1, then this condition is complex
, for the direct CP-violating asymmetry is very sensitive to the relative strong phase angle $\delta$, for example, $z_{a^{+}_0K^{*-}}=0.88$ (scenario I) and
$z_{a^{+}_0K^{*-}}=1.32$ (scenario II), though these two values are close to each other, but their corresponding direct CP-violating asymmetries
are very different.

In order to characterize the symmetry breaking effects and the contribution from tree operators and the electro-weak penguin,
it is useful to define the parameters below:
\be
R_1&=&\frac{{\cal B}(\bar B^0\to  a^{0}_0(1450)\bar K^{*0})}{{\cal B}(\bar B^0\to  a^{+}_0(1450)K^{*-})},\\
R_2&=&\frac{{\cal B}(B^-\to  a^{0}_0(1450)K^{*-})}{{\cal B}(B^-\to  a^{-}_0(1450)\bar K^{*0})},\\
R_3&=&\frac{\tau(B^0)}{\tau(B^-)}\frac{{\cal B}(B^-\to  a^{-}_0(1450)\bar K^{*0})}{{\cal B}(\bar B^0\to  a^{+}_0(1450)K^{*-})}.
\en
Considering the ratios of the branching ratios is a more transparent comparison between the predictions and the data because they
are less sensitive to the nonperturbative inputs. So the large deviation of these ratios from the standard-model predictions could
reveal a signal of new physics. When we ignore the tree diagrams and electro-weak penguins, $R_1, R_2$, and $R_3$ should be equal to $0.5, 0.5$, and
$1.0$. From our calculations, their values are:
\be
R_1=0.33, R_2=0.85, R_3=0.86,\mbox{ scenario I},\\
R_1=0.50, R_2=0.23, R_3=1.00,\mbox{ scenario II}.
\en
One can find the ratios $R_1$ and $R_3$ for scenario II are in agreement well with the predictions, while there is a large deviation for the
ratio $R_2$, and the reason is the aforementioned smallest penguin operator contributions for the channel $B^-\to  a^{0}_0(1450)K^{*-}$. For scenario
I, there are large deviation for all three ratios and the deviation for $R_2$ is the largest. These ratios can be tested by the future experiments.


\section{Conclusion}\label{summary}

In this paper, by using the decay constants and light-cone distribution amplitudes
derived from QCD sum-rule method, we calculate the branching ratios and the direct CP-violating
asymmetries of decays $B\to a_0(1450)K^*$
in the PQCD factorization approach and find that
\begin{itemize}
\item
For the decays $B^-\to  a^{-}_0(1450)K^{*0}, \bar B^0\to  a^{+}_0(1450)K^{*-}, a^{0}_0(1450)\bar K^{*0}$, their branching ratios in scenario II are
larger than those in scenario I about one order. So it is easy for the experiments to differentiate between the lowest lying state and the first
excited state for the meson $a_0(1450)$.
\item
For the decay $B^-\to  a^{0}_0(1450)K^{*-}$, due to not receiving the enhancement from the $K^*-$emission factorizable diagrams, its penguin operator
contributions are the smallest in scenario II, which makes its branching ratio drop into the order of $10^{-6}$, even so, its branching ratio in scenario
II is still larger than that in scenario I about $2.5$ times.

\item
The PQCD predictions are much larger than QCD factorization results. Because the latter can not make a reliable calculation from factorizable annihilation
diagrams and bring large uncertainties into the branching ratios, so they are still consistent with each other within the large theoretical errors.

\item
For these considered decays, their tree contributions are strongly CKM suppressed and they are penguin document decay modes. But the decay
$\bar B^0\to  a^{+}_0(1450)K^{*-}$ is abnormal in scenario I, for it receives an enhancement from the large Wilson coefficients $a_1=C_2+C_1/3$,
which makes its tree contribution survive the suppression.

\item
The direct CP-violating asymmetry is determined by the ratio of penguin to tree contributions, that is $z$. Generally speaking, if the value of $z$ is large, the
corresponding direct CP-violating asymmetry will be small, vice versa. While if the value of $z$ is very small and close to 1, the direct
CP-violating asymmetry will be sensitive to the relative strong phase angle $\delta$.
\end{itemize}

\section*{Acknowledgment}
This work is partly supported by the National Natural Science Foundation of China under Grant No. 11047158, and by Foundation of Henan University of Technology under Grant No.150374.

\end{document}